\documentstyle[psfig,12pt,twoside]{article}
\begin{document}
\pagenumbering {arabic}
\pagestyle{myheadings}
\markboth{J. Rafelski, J. Letessier and  A. Tounsi }
{Strange anti-baryons --- QGP versus HG}
\title{
          STRANGE ANTI-BARYONS \\ QGP VERSUS HG
}

\author{ \bf Johann Rafelski\thanks{In part supported by US-DOE 
grant DE-FG02-92ER40733} \\  
Department of Physics, University of Arizona\thanks{
Department of Physics, University of Arizona, Tucson, AZ 85721}\\$\ $\\
\bf Jean  Letessier \hspace{1.5cm} \bf Ahmed Tounsi\\  
Laboratoire de Physique Th\'eorique et Hautes Energies\thanks{ 
Unit\'e  associ\'ee au CNRS UA 280,
 Universit\'e PARIS 7, Tour 24, 
5\`e \'et., 2 Place Jussieu, F-75251 CEDEX 05.}  }
\date{}   
\maketitle  
\begin{abstract}
\noindent
We study quark-gluon plasma (QGP) and hadronic gas (HG) models of the
central fireball presumed to be the source of abundantly produced
strange (anti-)baryons in S $\to$ W collisions at 200 GeV A. We consider
how multi-strange (anti-)baryon multiplicities depend on strangeness
conservation and compare the HG and QGP fireball scenarios. We argue that
the total particle multiplicity emerging from the central rapidity region
as well as the variation of production rates with changes in the beam
energy allows to distinguish between the two reaction scenarios.
\end{abstract}

\vfill
\begin{center}
Published in:

 Proceedings of the XXVI International Conference on High
 Energy Physics

 Dallas, TX, August 1992

 American Institute of Physics, New York, J.R. Sanford, ed., pp. 983--990.
\end{center}
\vfill
{\bf AZPH--TH/92--27  \hfill}

\noindent{\bf PAR/LPTHE/92--35  \hfill September 1992}
\eject
\section{INTRODUCTION}

Kinetic strange particle production models \cite{KMR86} show that
abundant strangeness is suggestive of QGP formation in relativistic
nuclear collisions. Even more specific information about the nature of
the dense matter can be derived studying strange quark and anti-quark
clusters, which are more sensitive to the environment from which they
emerge \cite{RD87}, here in particular particle density. Therefore the
relative production abundances of strange and multi-strange  baryons and
anti-baryons where studied experimentally \cite{WA85}. They turned out to
be particularly sensitive probes of the thermal conditions of their
source \cite{MYSTR}. It has already been demonstrated that the observed
particle abundances are in agreement with a picture of explosively
disintegrating QGP fireball \cite{Raf91}.

It can be also argued \cite{CS92,LTR92} that these results are compatible
with the scenario of an equilibrium HG fireball, though here one lacks an
accepted mechanism for strangeness production. But perhaps more
importantly, it is hard to imagine a hadronic gas at temperature well
above the pion mass, {\it viz.} $T=215$ MeV which is the required
temperature for the hadronic gas interpretation of the strange anti-
baryon data to work \cite{LTR92}.

The problem we address here is how one can eliminate experimentally the
unnatural for the circumstances possibility of a hadronic gas fireball as
the underlying strange particle source. We suggested \cite{LTR92} that a
simple distinction of these two phases derives from the inherent
difference with regard to their entropy content $\cal S$ given a fixed
and conserved property, such as baryon number content  $\cal B$ which can
be  determined experimentally.  $\cal B$ is seen as being well understood
in terms of the nucleon number of the combined system of the projectile
nucleus and the target tube of nuclear matter cut out in the collision
from the much larger target nucleus. Baryon number of the fireball can
decrease only by particle radiation in the final disintegration of the
fireball, beyond which we assume that the scattering between the
different components have ceased and the relative abundances carry the
information about the property of the source.

On the other hand, once the pre-equilibrium reactions have been
terminated, and the particle momentum distributions have reached their
thermal form, entropy production effectively has ceased, even if a phase
transition occurs from a primordial phase to the final HG state. Hence
both baryon number and entropy content of the isolated fireball remain
constant and their ratio in a theoretical description is rather model
independent. Therefore a supplementary measurement, which will permit to
define the properties of this source is the multiplicity per
participating baryon in the fireball which is directly related to
entropy. While the hadronization of the entropy rich QGP fireball is
presently not understood, we take advantage here of the fact that in any
case a  substantially enhanced particle multiplicity must result, as
compared to the HG scenario. This can e.g. arise if the QGP fireball were
to evaporate emitting hadronic particles sequentially. For $T\sim215$ MeV
\cite{Raf92},  HG leads to about 40\% of the particle multiplicity
expected for QGP scenario \cite{LTR92}.

Alternatively, one may be able to distinguish the two cases  considering
the response of the measured parameters to changes in energy or/and size
of the colliding nuclei \cite{LTHR92}. We note that the hadronic gas
particle abundances are certain functions of the three thermal
parameters, $T$ the temperature, $\mu_{\rm B}$ the baryo-chemical
potential, and $\mu_{\rm s}$ the strange-chemical potential. The
constraint to a fixed strangeness fixes in HG a relation between
$\mu_{\rm B}$ and $\mu_{\rm s}$. Ideally, the number of strange and
antistrange quarks are equal, but pre-equilibrium emission can introduce
a small asymmetry. The strange-chemical potential $\mu_{\rm s}$ will in
case of QGP formation always remain independent of $\mu_{\rm B}$, and
near to the value $\mu_{\rm s}=0$, while the saturation of the phase
space described by a factor $\gamma$ (see below) will approach unity for
increasing size of the hot fireball. Neither result is generally correct
for the case of HG, and values of $\mu_{\rm s}$ and $\gamma$ have
considerable impact on the strange particle abundances.

To be able to forecast both for QGP and HG scenarios the behavior of
strange particle abundances as the nuclear collision energy changes we
study here how the strange-chemical potential $\mu_{\rm s}$ relates to
the baryo-chemical potential $\mu_{\rm B}$, for several choices of energy
density (temperature). We consider three bench mark temperatures: aside
of $T\simeq200$ MeV appropriate for the CERN--SPS energy range we also
include in our discussion $T\simeq 150$ appropriate for the lowest
conditions with probable QGP formation and $T\simeq300$ MeV, which we
judge appropriate for the BNL--RHIC facility under construction.

Our work \cite{LTR92,LTHR92} is in detail very different from the
parallel effort of Cleymans and Satz \cite{CS92,Cley92}. We study
particle ratios at fixed transverse mass $m_\bot>1.5$ GeV. We allow for
the strangeness phase space to be only partially saturated, and we
consider the degree of saturation to be experimentally measurable with
the result to be compared to the kinetic theory of strangeness
production. We allow the strangeness to be unbalanced (up to 10\%). We
compare the entropy contents for the different scenarios and confront the
findings with the observed particle multiplicity. We consider {S--S}
experimental results not to be in the same class as the S--W results (at
200 GeV A) due to the different stopping, and do not combine the data in
our analysis of the experimental results. We do not use kaon data, as
kaons, unlike strange anti-baryons can arise from peripheral, spectator
related processes and are therefore not necessarily witnesses of the same
stages of the collision.

\section{THERMAL FIREBALL MODEL}

We assume the formation in the collision of a region of space containing
much of the energy and baryon number available, with the hadronic
particles sharing the  accessible energy --- this we call a central
(rapidity) fireball. Our discussion is unaffected by the presence of a
collective flow, and the thermal parameters of the fireball are deduced
from the experimental results. In this analysis we imply that a rapid
disintegration of the fireball ensues its initial formation. Therefore it
is possible to use a simple average value of the thermal parameters for
the entire fireball neglecting the influence of flow on the spectra
\cite{Heinz92}. Recent studies of the dynamics of QGP to HG transition
\cite{CK92}  find that such a scenario is not impossible for the hot and
dense fireball we consider here.

For a hadronic fireball created in central S$\to$W collisions one has
about 108 baryons in the geometric interaction tube at small impact
parameter. In the HG fireball scenario we find that the strange pair
abundance is about 0.4 per baryon, given the thermal parameters
determined earlier \cite{LTR92}, corresponding to about 40 strange
particle pairs. In QGP fireball scenario the strangeness pair abundance
per baryon at $\gamma=0.7$ is about one, i.e., there is a strange
particle yield enhancement by about factor 2.5 as compared to hadronic
gas interactions.

\subsection{{\it u--d} asymmetry}

The ratio of the net number of down and up quarks in the fireball
\begin{eqnarray}
R_{\rm f}={\langle d-\bar d\rangle\over \langle u-\bar u\rangle}\ ,
\end{eqnarray}
arising in a S$\to$W-tube collisions is $R_{\rm f}^{\rm S-W}\simeq 1.08$
and in Pb--Pb collisions it is $R_{\rm f}^{\rm Pb-Pb}\simeq 1.15$. Taking
this into account we denote: $\mu_{\rm q}=(\mu_{\rm d}+\mu_{\rm u})/2$
and the asymmetry is related to $\delta\mu=\mu_{\rm d}-\mu_{\rm u}$. The
value of $\delta\mu$ is at each fixed $T$ given by the value of $R_{\rm
f}$, in dependence on the assumed structure of the source, i.e., the
equation of state. In the region of $T,\mu_{\rm B}$ of interest to us we
find that $\delta\mu/\mu_{\rm q}\sim R_{\rm f}-1$; though small, the
difference between the chemical potentials of $u$ and $d$ quarks is not
negligible.

\subsection{Partition function}

We have distinguished between the $u,d$ quarks in our calculations and
had computed the HG partition function distinguishing the flavor content
of strange and non-strange hadrons. We have summed explicitly the
contributions of hadronic particles with the mesons included up to mass
1690 MeV, nucleons up to 1675 MeV and $\Delta$'s up to 1900 MeV. Our
procedure will be evident when we discuss the strange particle sector
explicitly below. We note that higher hadronic resonances would matter
only if their number were divergent as is the case in the Bootstrap
approach of Hagedorn \cite{HAG78} and the HG was sufficiently long lived
to populate all high mass resonances. Our simple minded approach to
describe the HG phase is not sufficiently precise as soon as we ask
questions which are dependent either on the ever increasing mass spectrum
of particles or on the proper volume occupied by the particles
\cite{TOUNSI91}. However, quantities such as condition of zero
strangeness, fixed entropy per baryon are independent of the absolute
normalization of the volume and of the renormalization introduced by the
diverging spectrum and hence can be considered in the approach we take.

In order to simplify the comparison between our work and Ref.~\cite{CS92}
that while we always use $\mu_{\rm s}$ the strange (quark) chemical
potential, these authors use instead the chemical potential $\mu_{\rm S}$
of the kaons ($q\bar s$). The relation is: $\mu_{\rm s}=\mu_{\rm q}-
\mu_{\rm S}=\mu_{\rm B}/3-\mu_{\rm S}$. Here we employed the relation
between the baryo-chemical potential and quark chemical potential
$\mu_{\rm B}=3\mu_{\rm q}$ as follows from the baryon number carried by
quarks. In general it is not necessary to introduce different chemical
potentials (or fugacities) for the hadronic gas phase, as the chemical
potential of each HG species is simply the sum of the potentials of the
constituent quarks, {\it viz.}\/ for a proton $\mu_{\rm p}=2\mu_{\rm
u}+\mu_{\rm d}$, { etc}. Frequently, instead of chemical potentials the
fugacities $\lambda_i=\exp(\mu_i/T)$ are used.

In the Boltzmann approximation the partition function for the strange
particle fraction of the hadronic gas, ${\cal Z}_{\rm s}$ in the notation
of Ref.\,\cite{Raf87} is
\begin{eqnarray}
\ln{\cal Z}_{\rm s} = { {V T^3} \over {2\pi^2} }
\left\{(\lambda_{\rm s} \lambda_{\rm q}^{-1} +
\lambda_{\rm s}^{-1} \lambda_{\rm q}) \gamma F_K \right.\nonumber \\
\left.+(\lambda_{\rm s} \lambda_{\rm q}^{2} +
\lambda_{\rm s}^{-1} \lambda_{\rm q}^2) \gamma F_Y \right.\nonumber \\
\left.+ (\lambda_{\rm s}^2 \lambda_{\rm q} +
\lambda_{\rm s}^{-2} \lambda_{\rm q}^{-1}) \gamma^2 F_\Xi\right.\nonumber
\\ \left.+ (\lambda_{\rm s}^{3} + \lambda_{\rm s}^{-3})
\gamma^3F_\Omega\right\}\ , \label{4a}
\end{eqnarray}
where the kaon ($K$), hyperon ($Y$), cascade ($\Xi$) and omega ($\Omega$)
degrees of freedom in the hadronic gas are included successively.  Given
the recently measured values of the thermal parameters we are obliged to
sum over some more strange hadronic particles than was done in Eq.\,4 of
Ref.\,\cite{Raf87}. The phase space factors $F_i$ of the strange
particles are:
\begin{eqnarray}
F_K&=&\sum_j g_{K_j} W(m_{K_j}/T)
\ ,\nonumber\\
F_Y&=&\sum_j g_{Y_j} W(m_{Y_j}/T)
\ ,\nonumber\\
F_\Xi&=&\sum_j g_{\Xi_j} W(m_{\Xi_j}/T)
\ ,\nonumber\\
F_\Omega&=&\sum_j g_{\Omega_j} W(m_{\Omega_j}/T)
\ .
\label{FSTR}
\end{eqnarray}
We have included kaons up to 1650 MeV, hyperons up to 1750 MeV, cascades
up to 1820 MeV and also omegas up to 2250 MeV. We use  the notation
$W(x)=x^2K_2(x)$, and $K_2$ is the modified Bessel function. The factor
$\gamma$ in Eq.\,\ref{4a} allows us to consider the strange particle gas
away from absolute chemical equilibrium which corresponds to the value
$\gamma=1$. In general, if there is not sufficient time to make
strangeness (but sufficient time to exchange strange quarks between the
carriers, which we implicitly assumed above) the partition function
applies with $\gamma<1$. The value of the factor $\gamma$ is determined
by the dynamics of strangeness production. Its measurement is only
possible in the comparison of abundances of hadrons comprising different
numbers of strange (or antistrange) quarks. The value $\gamma=0.7\pm0.1$
\cite{MYSTR} arising from  the WA 85 results \cite{WA85} is suggestive of
QGP based  strangeness production mechanisms.

\subsection{Strangeness balance}

We consider a HG fireball in which the number of $s$ and $\bar s$ quarks
is (nearly) equal. The condition that the total strangeness vanishes
takes the form
\begin{equation}
0=\langle s \rangle - \langle \bar s \rangle=
\lambda_{\rm s} { \partial \over {\partial \lambda_{\rm s}}}
\ln {\cal Z}_{\rm s}\ .
\label{Eq6}
\end{equation}
This is an implicit equation relating $\lambda_{\rm s}$ with
$\lambda_{\rm q}$ for each given $T$. A slight generalization is obtained
allowing that a small fraction imbalance in strange quark numbers arising
from pre-equilibrium emission \cite{LTHR92} and effects of up to 10\%
were considered.

\section{CONSTRAINT BETWEEN $\mu_{\rm s}$ AND $\mu_{\rm B}$}

In order to better understand the numerical results we first study an
aspect of the condition  (\ref{Eq6}) analytically. We seek for $\mu_{\rm
s}=0$, {\it viz.}\/ $\lambda_{\rm s}=1$, non trivial values of $\mu_{\rm
B}^0$ (different from the trivial solution $\mu_{\rm B}^0=0$) for which
strangeness balances out. We find the exact answer to be:
\begin{eqnarray}
\mu_{\rm B}^0=3{\rm cosh}^{-1}\left({F_{\rm K}\over 2F_{\rm Y}}
-\gamma {F_{\Xi}\over F_Y}\right)\ .
\end{eqnarray}
There is a real solution only when the argument on the right hand side is
greater than unity. It turns out that this condition is a sensitive
function of the temperature $T$ and of the hadronic resonances included
in Eq.\,\ref{FSTR}. For any given spectrum used to compute the phase
space factors $F_i$ there is a temperature $T_0$ beyond which no such
solution is possible --- this occurs since $F_{\rm K}/F_{\rm Y}$ is a
monotonically decreasing function of $T$ (see Fig.\,1,
Ref.\,\cite{Raf87}).

\begin{figure}[t]
\vspace{ 0cm}
\centerline{\hspace{8cm}\psfig{figure=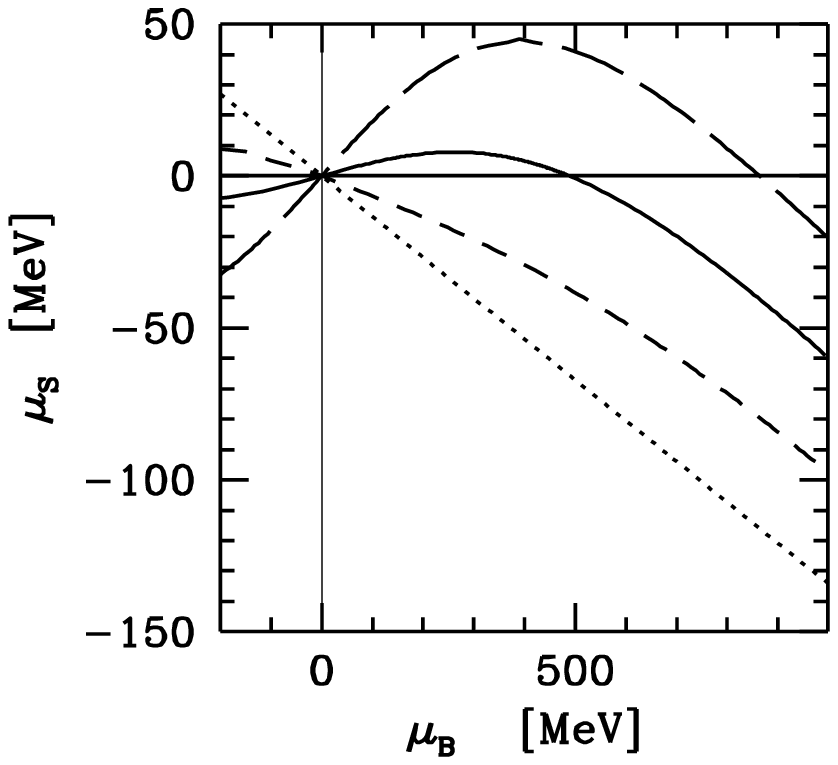,height=22cm}}
\vspace{-14cm}
\vbox{\parindent=0pt\baselineskip=12pt\small Figure~1.
Strange-chemical potential $\small\mu_{\rm s}$ versus baryo-chemical
potential $\small\mu_{\rm B}$ for zero strangeness fireball. Long-dashed
line corresponds to $T= 150$~MeV, solid line to $ T = 200$~MeV, and
dashed line to $T = 300$ MeV. The dotted line is the limiting curve for
large~$T$.}
\label{F1}
\end{figure}

In consequence of the above observations we expect that there is a domain
of temperatures for which even a considerable change of $\mu_{\rm B}$
does not induce a significant change of $\mu_{\rm s}$. For the complete
set of resonances we have included in the phase space factor, this quite
peculiar behavior occurs just at the temperature $T\simeq215$ MeV. In
Fig.\,\ref{F1} we present the constraint between $\mu_{\rm s}$ and
$\mu_{\rm B}$ at fixed $T=200$ MeV (solid curves), 150 MeV (long-dashed
curves), 300 MeV (short-dashed curves) and 1,000 ($\simeq\infty$) MeV
(dotted curves) for $\gamma=0.7$ (the choice $\gamma=1$ influences this
result insignificantly).
The solid lines corresponding to $T=200$ MeV is indeed leading to quite
small values of $\mu_{\rm s}$ for all $\mu_{\rm B}<500$ MeV.

\subsection{Strange baryon ratios}

This behavior explains why in the vicinity of $T=215\pm15$ MeV the QGP
and HG are leading to the same particle abundances: the constrain to zero
strangeness in HG is consistent with $\mu_{\rm s}$ characteristic of QGP
phase. We also find as Fig.\,\ref{F1} clearly shows, that this ambiguity
does not persist at higher or lower temperature. To make this point more
quantitative, we now study strange baryon ratios arising from a thermal
fireball {\em constrained} to vanishing strangeness.

\begin{figure}[t]
\vspace{0cm}
\centerline{\hspace{8cm}\psfig{figure=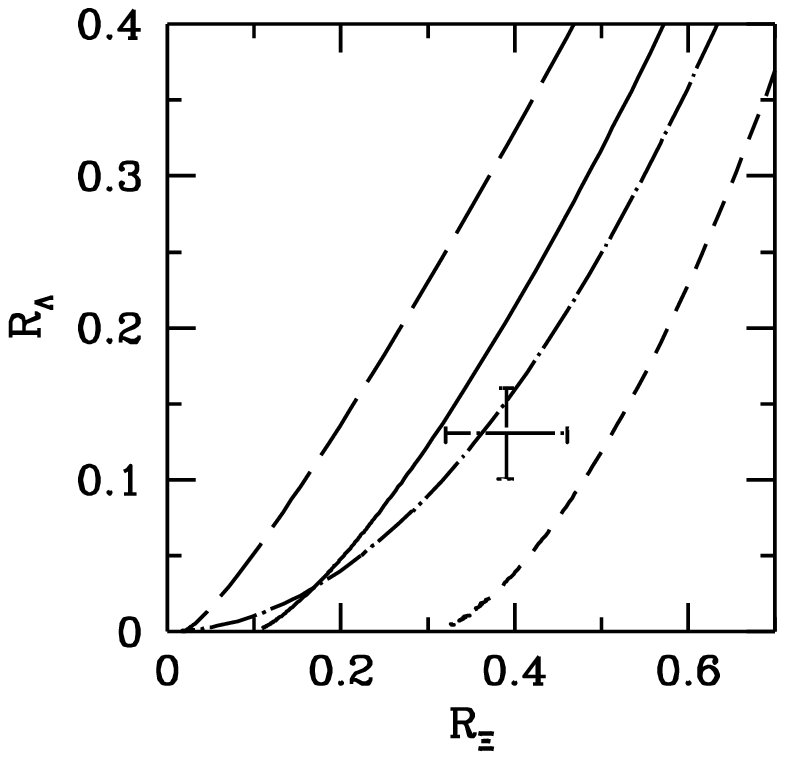,height=22cm}}
\vspace{-14cm}
\vbox{\parindent=0pt\baselineskip=12pt\small Figure~2.
$R_\Lambda$ versus $R_\Xi$.  Long-dashed line {corresponds} to $ T=
150$~MeV,  solid line to $ T = 200$~MeV, and dashed line to $T = 300$~MeV
in HG. The dashed-dotted  line corresponds to QGP.}
\label{F2}
\end{figure}

Comparing {\it spectra\/} of particles with their antiparticles within
overlapping regions of $m_\bot$, the Boltzmann and {\em all} other
statistical factors cancel, and their respective abundances are only
functions of fugacities \cite{RD87}. In Fig.\,\ref{F2} we show for the
case of exactly vanishing strangeness the resulting relation of
$R_\Xi={{\overline{\Xi^-}}/ {\Xi^-}}$ with
$R_\Lambda={\overline{\Lambda}/ \Lambda}$. In addition to the HG results
for temperatures $T=200$ MeV (solid line), $T=150$ MeV (dashed line) and
$T=300$ MeV (dotted line) we show the case $\mu_{\rm s}=0$ corresponding
to QGP source (dashed-dotted line). The cross corresponds to the result
reported by the WA85 experiment \cite{WA85}. As can be seen, the QGP
curve will nearly coincide with the $T=215$ MeV curve in the HG case, as
we noted before \cite{LTR92}.

In addition to the baryon ratios one can also consider the ratio of kaons
to hyperons, again at fixed $m_\bot$. Because of the experimental
procedures used, which rely on the observation of the disintegration of
neutral strange particles into two charged products a comparison of the
$K_{\rm s}$ (here {\em s} stands for {\em short\/}) with the $\Lambda$
(which includes the $\Sigma^0$ abundance) is available. Considering
$R_{\rm K}={K_{\rm s}/( \Lambda+\Sigma^0)}$ as function of $R_\Lambda$ we
find that there is poor sensitivity of the result to the nature of the
fireball, $R_{\rm K}$ is a good measure of the baryo-chemical potential
\cite{KMR86} of the source of these particles, which may be in part the
fragmentation region of the heavy target nucleus.

\subsection{Measurement of $\gamma$}

We have introduced the factor $\gamma$ which characterizes the approach
to saturation density of the strange quark abundance. It can be
experimentally measured by determining the product:
\begin{equation}
{{\Xi\cdot\overline{\Xi}}\over{\Lambda\cdot\overline{\Lambda}}}=\gamma^2\
. \end{equation}
It is easy to see that all spectral and chemical factors cancel in this
combination of particle abundances, and the only unbalanced factor is the
phase space density of strangeness. We refer to reference \cite{LTHR92}
for further details.

\section{DISCUSSION}

\begin{figure}[t]
\vspace{0cm}
\centerline{\hspace{8cm}\psfig{figure=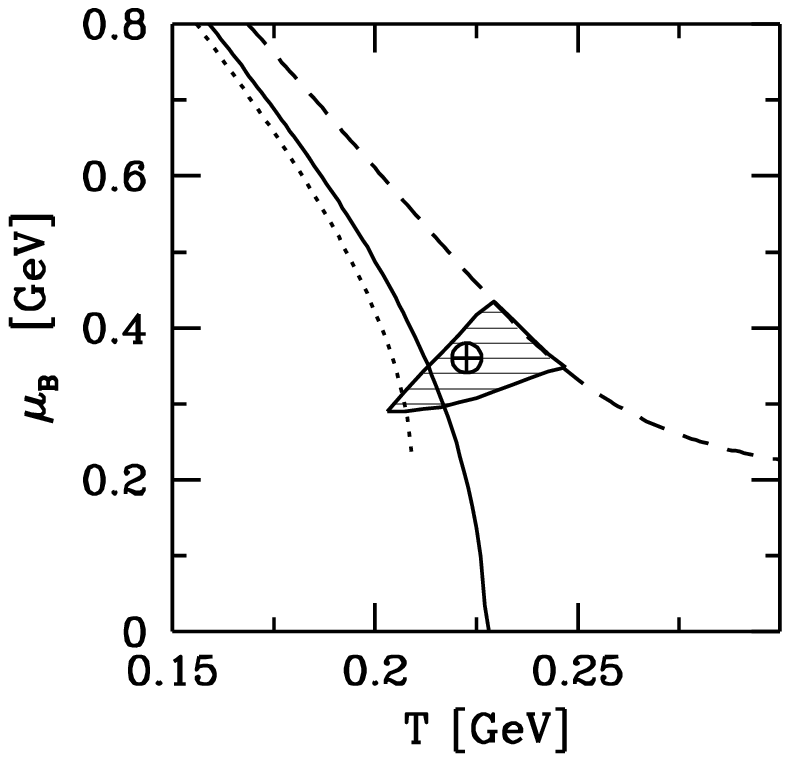,height=22cm}}
\vspace{-14cm}
\vbox{\parindent=0pt\baselineskip=12pt\small Figure~3.
Solid line shows in the $\small\mu_{\rm B}$--$T$ plane the condition of
zero strangeness in HG fireball assuming the QGP-like condition
$\small\lambda_{\rm s}=1$; dashed line $\small\lambda_{\rm s}= 0.95$;
dotted line $\small\lambda_{\rm s}=1.02$. Hatched is the region
compatible with the experimental WA85 data. The $\oplus$ corresponds to
$T =220$~MeV, $\small\mu_{\rm B}=340$~MeV, the central point for QGP
fireball.}
\label{F3}
\end{figure}

In the $\mu_{\rm B}$--$T$ plane the condition of zero strangeness
combined with the condition $\lambda_{\rm s}=$~1 leads to the curve shown
in Fig.\,\ref{F3} by the solid line. Dashed line is the case
$\lambda_{\rm s}=0.95$ and dotted line $\lambda_{\rm s}=1.02$. The upper
and lower boundary of hatched area arise from $\mu_{\rm
B}/T=3\cdot0.52\pm0.01$ and from the constraint obtained from  the $K^-
/\Lambda$ ratio reported \cite{WA85}.  We find that if we are willing to
accept a hadronic gas \cite{CS92} at temperature of $T\simeq 200-210$
MeV, it could indeed be the source of strange particles --- a puzzle in
such an interpretation is the condition of $\lambda_{\rm s}\simeq 1$
which is natural for QGP, and does not have at present any special
founding for the HG state.

\section{Particle Multiplicity}

The properties of the HG and QGP fireballs are considerably different in
particular with regard to the entropy content. Both states are easily
distinguishable in the regime of values $\mu_{\rm B},\ T$ shown in
Fig.\,\ref{F3}. We find for the entropy per baryon ${\cal S}^{\rm
HG}/{\cal B} =21.5\pm1.5$. Consequently, the pion multiplicity which can
be expected from such a HG fireball is $4\pm0.5$. This is less than half
of the QGP based expectations we found in \cite{LTR92}, and clearly the
difference is considerable in terms of experimental sensitivity. Checking
the theoretical sensitivity we find that the point at which the entropy
of HG and QGP coincide {\it and} strangeness vanishes {\it and}
$\lambda_{\rm s}\simeq1$ is at $T\simeq135$ MeV, $\mu_{\rm B}\simeq950$
MeV, quite different from the region of interest here. We note that
charged particle multiplicity {\it above} 600 in the central region has
been seen \cite{mult} in heavy ion collisions corresponding  possibly to
a total particle multiplicity of about 1,000, as required in the QGP
scenario for the central fireball we described above.

\begin{figure}[t]
\vspace{0cm}
\centerline{\hspace{8cm}\psfig{figure=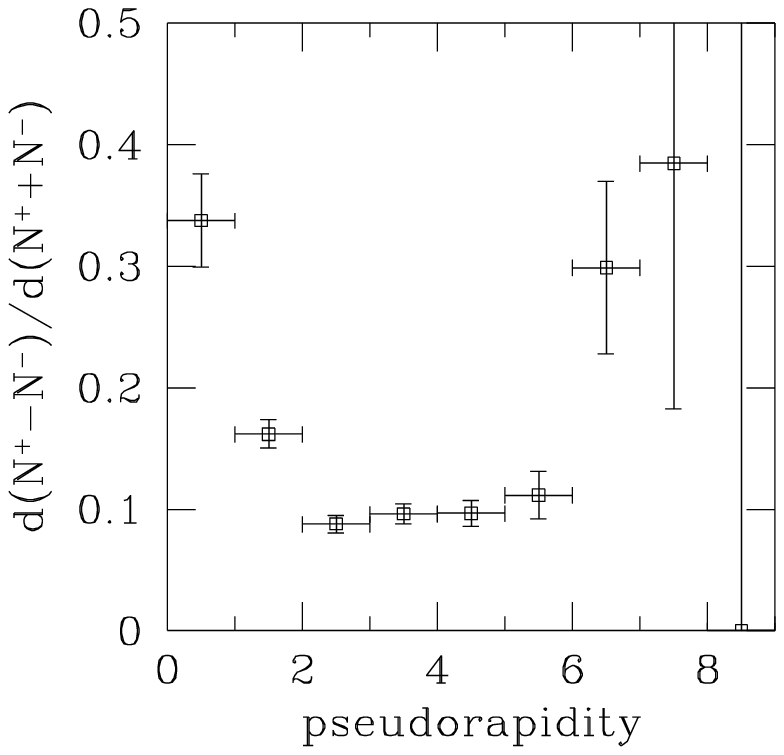,height=22cm}}
\vspace{-14cm}
\vbox{\parindent=0pt\baselineskip=12pt\small Figure~4.
Emulsion data for charged particle multiplicity as function of pseudo-
rapidity:  difference of positively and negatively charged particles
normalized by sum of both polarities. (Courtesy of CERN-EMU 05
collaboration, Y. Takahashi et al. \cite{mult}).}
\label{F4}
\end{figure}

Some of these emulsion particle multiplicity data are shown as function
of rapidity in Fig.\,\ref{F4}. Here we draw \cite{mult} $D_{Q}$, the
difference in the number of positively and negatively charged particles
normalized by their sum. All up to date scanned (15) events of 200 GeV A
S-Ag interactions with the ``central" trigger being the requirement for
the total charged multiplicity $> 300$. Reaction spectators (target
fragments) are not observed in this experiment. In absence of strange
particles we find assuming pion symmetry $\pi^+=\pi^-=\pi^0=\pi/3$:
\begin{equation}
D_Q\equiv{{N^+-N^-}\over{N^++N^-}} \to
{{\cal B}\over\pi}\ {1.5\over{1+R_{\rm f}+1.5\,{\cal B}/\pi}}
\end{equation}
At central rapidity a value of 0.08--0.09 is found. For a hadronic gas
with small strange particle component we would have expected based on the
entropy argument a value more than twice as large. In a numeric
calculation in which we take $\lambda_{\rm s}=1\pm0.05$ and fix the
temperature for each $\mu_{\rm B}$ such that strangeness is conserved, we
find:
\begin{equation}
D_Q = {\mu_{\rm B}\over\mbox{1.3 GeV}}\ \mbox{for } \mu_{\rm B}<0.6
\mbox{ GeV}.
\label{DQ}
\end{equation}
This HG result is extremely simple, considering the complexity of the
calculation.  We thus see that in the HG scenario for the strangeness
source which {\em has to have} $\mu_{\rm B}\sim 0.35$ GeV, see
Fig.\,\ref{F3}, Eq.\,\ref{DQ} implies $D_Q\sim 0.27$, which is
incompatible with the EMU 05 data \cite{mult}, as is seen in
Fig.\,\ref{F4}. Along with this observation that HG is incompatible with
he combined EMU 05 and WA85 data goes the fact we discussed here at
length that the characteristic property of QGP is the persistence of the
fireball parameters $\mu_{\rm s}=0,\ \gamma\sim1$ irrespective of the
temperature reached.

\subsection{Dependence on Temperature}

Only for a temperature in the vicinity of 200--215 MeV the HG scenario
also leads to the value $\mu_{\rm s}=0$, while there is no ready
explanation how $\gamma\sim1$ is reached.  We have identified the reason
for this coincidence to be the peculiar behavior of $\mu_{\rm s}$ as
function of $\mu_{\rm B}$ when the constrain on the total fireball
strangeness is imposed. Thus it is quite natural to expect that at the
lower energy accessible to CERN--SPS (60 GeV S-beam and about 50 GeV Pb-
beam) a lower value of temperature accompanied by higher value of
$\mu_{\rm B}$ will be reached.  We stress that {\em if the strange
particles emerge from nearly zero strangeness fireball\/} formed at a
given temperature (150, 200, 300 MeV) it is imperative that the observed
particle ratios fall on the here presented lines of Fig.\,\ref{F2},

In conclusion we note that the currently available strange particle
spectra, though consistent both with QGP and/or HG at $T>200$ MeV, are
strongly favoring the QGP interpretation (because of $\lambda_s=1,\
\gamma\simeq 1$). This case is considerably strengthened by the
consideration of the emulsion multiplicity data. Our discussion therefore
strongly suggests that strange anti-baryon abundances be measured along
with non-strange particle multiplicities and that it is desirable to
widen the study of strange anti-baryon ratios to higher {\em and\/} lower
nuclear beam energies.


\end{document}